\documentclass[11pt]{article}
\usepackage{amsmath,amsfonts,enumitem}
\def \beq  {\begin{equation}}
\def \eeq  {\end{equation}}
\def \beqar {\begin{eqnarray}}
\def \eeqar {\end{eqnarray}}
\def\sqr#1#2{{\vcenter{\vbox{\hrule height.#2pt
\hbox{\vrule width.#2pt height#1pt \kern#1pt
\vrule width.#2pt}\hrule height.#2pt}}}}

\def\S {{\cal S}}

\def\Tr {{\rm Tr}}

\def\del {\partial}

\def\half{\textstyle{1\over 2}}

\begin{document}
%%%%%%%%%%%%%%%%%%%%%%%%%%%%%%%%%%%%%%%%%%%%%%%
%\fontfamily{pnb}\fontsize{12pt}{16pt}\selectfont
%\fontfamily{pzc}\fontsize{14pt}{16pt}\selectfont
%\fontfamily{pbk}\fontsize{12pt}{16pt}\selectfont
\fontfamily{cmr}\fontsize{11pt}{14pt}\selectfont
%\fontfamily{phv}\fontshape{ro}\fontsize{11pt}{14pt}\selectfont
%\fontfamily{put}\fontshape{ro}\fontsize{11pt}{14pt}\selectfont
%\fontfamily{pmn}\fontshape{ro}\fontsize{11pt}{14pt}\selectfont
%\fontfamily{ptm}\fontseries{m}\fontshape{r}\fontsize{12pt}{16pt}\selectfont
%\fontfamily{pnc}\fontseries{m}\fontshape{r}\fontsize{11pt}{15pt}\selectfont
%\fontfamily{ppl}\fontseries{m}\fontshape{r}\fontsize{11pt}{15pt}\selectfont
%\usefont{T1}{phv}{m}{it}
%%%%%%%%%%%%%%%%%%%%%%%%%%%%%%%%%%%%%%%%%%%%%%%
\def \CMP {{Commun. Math. Phys.}}
\def \PRL {{Phys. Rev. Lett.}}
\def \PL {{Phys. Lett.}}
\def \NPBProc {{Nucl. Phys. B (Proc. Suppl.)}}
\def \NP {{Nucl. Phys.}}
\def \RMP {{Rev. Mod. Phys.}}
\def \JGP {{J. Geom. Phys.}}
\def \CQG {{Class. Quant. Grav.}}
\def \MPL {{Mod. Phys. Lett.}}
\def \IJMP {{ Int. J. Mod. Phys.}}
\def \JHEP {{JHEP}}
\def \PR {{Phys. Rev.}}
\def \JMP {{J. Math. Phys.}}
\def \GRG{{Gen. Rel. Grav.}}
%%%%%%%%%%%%%%%%%%%%%%%%%%%%%%%%%%%%%%%%%%%%%%%
%%%%%%%%%%%%%%%%%%%%%%%%%%%%%%%%%%%%%%%%%%%%%%%
\begin{titlepage}
\null\vspace{-62pt} \pagestyle{empty}
\begin{center}
\rightline{CCNY-HEP-11/7}
\rightline{December 2011}
\vspace{1truein} {\Large\bfseries
Fluids, Anomalies and the Chiral Magnetic Effect:}\\
\vskip .1in
{\Large\bfseries
A Group-theoretic Formulation}\\
\vspace{6pt}
\vskip .1in
{\Large \bfseries  ~}\\
\vskip .1in
{\Large\bfseries ~}\\
%%%%%%%%%%%%%%%%%%%%%%%%%%%%%%%%%%%%%%%%%%%%%%%%\vspace{.6in}
{\large\sc V.P. Nair$^a$, Rashmi Ray$^{a,b}$ {\rm and}  Shubho Roy$^{a,c}$}\\
\vskip .2in
{\itshape $^a$Physics Department\\
City College of the CUNY\\
New York, NY 10031}\\
\vskip .1in{\itshape $^b$Physical Review D\\
American Physical Society\\ 
Ridge, NY 11367}\\
\vskip .1in{\it $^c$ Department of Physics and Astronomy\\
Lehman College of the CUNY\\
Bronx, NY 10468}\\
\vskip .1in
\begin{tabular}{r l}
E-mail:
&{\fontfamily{cmtt}\fontsize{11pt}{15pt}\selectfont vpn@sci.ccny.cuny.edu}\\
&{\fontfamily{cmtt}\fontsize{11pt}{15pt}\selectfont ray@aps.org}\\
&{\fontfamily{cmtt}\fontsize{11pt}{15pt}\selectfont sroy@ccny.cuny.edu}
\end{tabular}

\fontfamily{cmr}\fontsize{11pt}{15pt}\selectfont
\vspace{.8in}
%\vspace{1.5in}%\vspace{0.3in}
%%%%%%%%%%%%%%%%%%%%%%%%%%%%%%%%%%%%%%%%%%%%%%%%%%%%%%%%%%%%
\centerline{\large\bf Abstract}
\end{center}
It is possible to formulate fluid dynamics in terms of group-valued variables.
This is particularly suited to the cases where the fluid has nonabelian charges and is coupled to nonabelian gauge fields. We explore this formulation further in this paper. An action for a fluid of relativistic particles (with and without spin) is given in terms of the Lorentz and Poincar\'e (or de Sitter) groups. Considering the case of particles with flavor symmetries, a general fluid action which also incorporates all flavor anomalies is given. The chiral magnetic and chiral vorticity
effects as well as the consequences of the mixed gauge-gravity anomaly are discussed.

\end{titlepage}

%%%%%%%%%%%%%%%%%%%%%%%%%%%%%%%%%%%%%%%%%%%%%%%%%%%%%%
\pagestyle{plain} \setcounter{page}{2}

\section{Introduction}

The description of fluid dynamics, especially for systems made of particles carrying nonabelian charges, has become an important research topic with the discovery of
of the state of unconfined quark and gluons, the quark-gluon plasma, in heavy ion collisions.
Field theoretic analyses, augmented with Boltzmann-type kinetic equations, can be used to
``derive" the equations of fluid dynamics, but are generally limited to dilute systems near equilibrium.
However, the basic equations can be formulated using general principles and therefore have a regime of validity significantly beyond the context of the derivation based on kinetic equations.
The question of a derivation based on symmetry principles generalizing
the usual equations of magnetohydrodynamics to include nonabelian charges and fields
is interesting in its own right, but has 
enhanced relevance after the discovery of the quark-gluon plasma.
Such an approach was developed in \cite{Bistrovic:2002jx, review}, where the fluid degrees of freedom were shown to be naturally described by the elements of a Lie group.
The method applies to ordinary hydrodynamics as well, but becomes particularly useful for
incorporating nonabelian symmetries and coupling to nonabelian fields.

A new impetus to such analyses has come from the recent work on the chiral magnetic effect
\cite{Kharzeev:2004ey}. 
The specific case of interest has been the charge separation and a corresponding electromagnetic current induced by the axial anomaly, which can be demonstrated by
the standard diagrammatic techniques. An interesting question to ask is then:
Is there an effective description of the anomalies and how they affect the fluid dynamics?
A related question is one of generalization to all flavor anomalies, even though
they may not be of immediate relevance to the quark-gluon plasma.
Symmetries are obviously front-and-center in analyzing anomalies and so our approach to fluid dynamics based on group-valued variables would seem tailor-made for these questions.
This is exactly the subject of the present paper.

The chiral magnetic effect, we may note, has led to a significant body of
 literature on related topics.
The possibility of describing the chiral magnetic effect using hydrodynamics and thermodynamics 
is explored in \cite{arXiv:1010.1550}. Transport in arbitrary dimensions induced by anomalies has also been discussed
in \cite{loganayagam}. Since there is considerable evidence that the quark-gluon plasma can be described as a strongly coupled fluid, the holographic correspondence can provide 
another method towards its analysis. The description of the chiral magnetic effect
using holographic approaches such as the AdS/CFT correspondence or the Sakai-Sugimoto model 
is given in \cite{arXiv:0904.4772}.
There is also an attempt to understand the chiral magnetic effect \cite{arXiv:1102.4334}
using the fluid/gravity correspondence of \cite{arXiv:0712.2456}.
For completeness, we also note that there have been many lattice simulations of the chiral magnetic effect \cite{arXiv:0907.0494}.

The focus in most of the literature has been on the computation of transport coefficients or the modifications of the energy-momentum tensor and the currents, and then the subsequent incorporation of these in the equations of motion of fluid dynamics.
 Our approach will be to write an effective action for anomalies directly in the fluid language, in other words, we obtain the fluid version of the Wess-Zumino term for anomalies.
 The action-based approach gives a simple starting point for all flavor anomalies. We also discuss some aspects of the mixed gauge-gravity anomaly in the standard model.
(The mixed anomalies, and the possibility of a chiral vortex effect, are also considered from the point of view of Kubo formulas and transport coefficients in \cite{landsteiner}.)
 On the negative side, the action-based approach will not include dissipative effects; they have to be added on after the equations of motion have been obtained by the variational principle.
We note that an effective action approach has been given in two dimensions \cite{nicolis}, although
the formulation is still very different from ours.

In section 2, we give a brief resume of the formulation of ordinary,
nonrelativistic or relativistic, fluid dynamics in terms of group variables.
We then describe how nonabelian internal symmetries are included to obtain a nonabelian
magnetohydrodynamics. In section 3, we follow a similar approach to construct the action for a
fluid of spinning particles in terms of the Lorentz group or Poincar\'e (or de Sitter) groups, the latter being adaptable to the spinless case as well. The fluid description for the quarks in the standard model is given in section 4, taking a fluid of the up, down and strange quarks as an example.
The full fluid action for these degrees of freedom including anomalies is given in this section.
The standard chiral magnetic effect, the chiral vorticity effect and mixed gauge-gravity anomalies are discussed in this framework in
section 5.

\section{Lagrangians and perfect fluids: a short resume}

We start with a recapitulation of the formulation of hydrodynamics
in terms of group theory. We will be brief, since this is reviewed
in detail in \cite{{Bistrovic:2002jx}, {review}}. Ordinary fluid dynamics can
be viewed as a Poisson bracket system with
\begin{equation}
\left[F,G\right]=\int\left[\frac{\delta F}{\delta\rho}\partial_{i}\left(\frac{\delta G}{\delta v_{i}}\right)-\frac{\delta G}{\delta\rho}\partial_{i}\left(\frac{\delta F}{\delta v_{i}}\right)-\omega_{ij}\frac{\delta F}{\delta v_{i}}\frac{\delta G}{\delta v_{j}}\right]\label{eq: hydrodynamics poisson bracket}
\end{equation}
for $F,G$ which are functions of the density $\rho$ and fluid velocity
$v_{i}$. The Hamiltonian,
\begin{equation}
H=\int d^{3}{\bf x}\left[\frac{1}{2}\,\rho\,{\bf v}^{2}+V(\rho)\right]\label{eq:fluid Hamiltonian}
\end{equation}
is then easily verified, via the brackets (\ref{eq: hydrodynamics poisson bracket})
to lead to the continuity and Euler equations, the fluid pressure
being $P=\rho\frac{\partial V}{\partial\rho}-V.$

The difficulty with this framework is also well known. The helicity
$C$, defined by,
\begin{equation}
C=\frac{1}{8\pi}\int\epsilon^{ijk}\, v_{i}\, \partial_{j}v_{k}\label{eq: fluid helicity}
\end{equation}
is seen to Poisson commute with all observables, i.e. $\left[F,C\right]=0$
for all $F$. Viewing this from a quantum point of view, we see that
the values of $C$ are superselected. It is therefore necessary to
specify a value for $C$ and consider the restricted Hamiltonian dynamics
for that sector itself. Alternatively, if we think of the Poisson
brackets to be written as $\left[\xi_{a},\xi_{b}\right]=K_{ab}$,
for $\xi$ being $\rho$ and $v_{i}$,, then the symplectic structure
is obtained as the inverse of $K_{ab}$. (Usually in starting from
a Lagrangian, we obtain the symplectic structure and invert it to
obtain the Poisson brackets.). Since $C$ commutes with all observables,
we see that $({\delta C}/{\delta v_{i}})$ is a zero mode for $K$
and hence we cannot relate Eqs. (\ref{eq: hydrodynamics poisson bracket},\ref{eq:fluid Hamiltonian})
to a symplectic structure or Lagrangian description without first
restricting the value of $C$. Thus, to obtain a Lagrangian description
we must first fix $C$ and then seek a parametrization for $v_{i}$
which does not further change the value of $C$. This is given by
the Clebsch parametrization,
\begin{equation}
v_{i}=\partial_{i}\theta+\alpha\, \partial_{i}\beta\label{eq:Clebsch parametrization}
\end{equation}
for an arbitrary functions $\theta,\alpha,\beta$. In this case $C=0$
if $\theta,\alpha,\beta$ are single-valued nonsingular functions
vanishing at infinity. (They are known as the Monge potentials.)
A suitable action for fluid dynamics is then \cite{review, lin}
\begin{equation}
S=\int\rho\,\dot{\theta}+\rho\,\alpha\,\dot{\beta}-\left[\frac{1}{2}\,\rho\,{\bf v}^{2}-V\right].\label{eq:fluid action}
\end{equation}
We note that $(\rho,\theta)$, $(\rho\alpha,\beta)$ form two sets
of canonically conjugate pairs.

Now we introduce an element $g$ of the group $SU(1,1)$ which may
be parametrized in general as,
\begin{equation}
g=\frac{1}{\sqrt{1-\bar{u}u}}\left[\begin{array}{cc}
1 & u\\
\bar{u} & 1
\end{array}\right]\left[\begin{array}{cc}
e^{i\theta/2} & 0\\
0 & e^{-i\theta/2}
\end{array}\right]\label{eq:SU(1,1) parametrization}
\end{equation}
where $u$ is a complex variable. Direct calculation shows that
\begin{equation}
-i \Tr \left(\sigma_{3}\,g^{-1}\,dg\right)=d\theta+\alpha \,d\beta,\label{eq: SU(1,1) Clebsch}
\end{equation}
\[
\alpha=\frac{\bar{u}u}{1-\bar{u}u},\:\beta=\left(-i/2\right)\ln\left(\frac{u}{\bar{u}}\right).
\]
The $\theta$-direction in $g$ corresponds to the compact direction,
the $U(1)$ subgroup generated by ${\half} \sigma_3$, while $\alpha$
and $\beta$ parametrize $SU(1,1)/U(1)$. The action (\ref{eq:fluid action})
can now be written as,
\begin{equation}
S=-i\int j^{\mu}\, \Tr\left(\sigma_{3}g^{-1}\partial_{\mu}g\right)-\int\left[\frac{j_{i}j_{i}}{2\rho}+V\right]\label{eq:group theoretic fluid action}
\end{equation}
where we denote $j^{0}=\rho$. The elimination of $j_{i}$ in (\ref{eq:group theoretic fluid action})
leads to the version ($\ref{eq:fluid action}$).

The relativistic generalization of fluid dynamics and the action (\ref{eq:group theoretic fluid action})
is also very straightforward. It is given by
\begin{equation}
S=-i\int j^{\mu}\, \Tr\left(\sigma_{3}g^{-1}\partial_{\mu}g\right)-F(n)\label{eq:group theoretic action for relativistic fluid}
\end{equation}
where $F(n)$ is a function of the variable $n$, which is defined
by $j^{\mu}j_{\mu}=n^{2}$. Equivalently
\begin{equation}
j^{\mu}=n\, u^{\mu}\label{eq:relativistic fluid current}
\end{equation}
where $u^{\mu}$ is a four-vector obeying $u^{\mu}u_{\mu}=1$. It
may be considered as the four-velocity of the fluid and $n$ identified
as the invariant density. The energy-momentum tensor for (\ref{eq:group theoretic action for relativistic fluid})
has the perfect fluid form
\begin{equation}
T^{\mu\nu}=nF'u^{\mu}u^{\nu}-g^{\mu\nu}(nF'-F)\label{eq:stress tensor of relativistic fluid}
\end{equation}
identifying the pressure as $P=nF'-F$. The function $F$ is thus
the enthalpy.

We have obtained a group theoretic formulation of ordinary fluid dynamics.
The existence of a compact direction, namely, the $U(1)$ direction
of the $SU(1,1)$, may seem a little puzzling at first, since at the
level of the classical Clebsch parametrization, this was not a requirement.
The Poisson bracket obtained from (\ref{eq:group theoretic fluid action})
or (\ref{eq:group theoretic action for relativistic fluid}) gives
\begin{equation}
\left[\rho(f),g(x)\right]=-ig(x)\frac{\sigma_{3}}{2}f(x).\label{eq:SU(1,1) PB}
\end{equation}
This means that in the quantum theory
\begin{equation}
U^{\dagger}\,g\,U=g\, e^{i\pi\sigma_{3}}=-g\label{eq:half pi sign flip}
\end{equation}
for $U=\exp\left[-2\pi i\int\rho\right]$. Since all observables involve
even powers of $g$, they are invariant under the action of $U$.
This
means that we can set $U=1$, giving $\int\rho=N$ for some integer
$N$. The existence of the compact direction thus requires the quantization
of $\int\rho$ in the quantum theory; this is equivalent to saying
that the fluid is made of particles with $\rho$ being the particle
density \cite{nair-ray}. Thus, rather than a defect of the group-theoretic parametrization
(\ref{eq: SU(1,1) Clebsch}) in comparison to the classical Clebsch
parametrization (\ref{eq:Clebsch parametrization}), we view this
as a good feature of the description in (\ref{eq:group theoretic fluid action}),
(\ref{eq:group theoretic action for relativistic fluid}). {[}If vorticity
were also quantized we would use $SU(2)$ in place of $SU(1,1)$.{]}

It is now easy enough to obtain the generalization to carrying nonabelian
gauge charges, corresponding to a compact Lie group $G$.

First consider $SU(2)$. At the particle level, the dynamics of a
particle carrying $SU(2)$ charges is given by the Wong equations
which have the action \cite{{wong},{bal1}}
\begin{equation}
S=\int\left[\frac{1}{2}\,m\,\dot{{\bf x}}^{2}+A_{i}^{a}Q^{a}\dot{x}_{i}-i\,w\,\Tr(\sigma_{3}g^{-1}\,\dot{g})\right]\label{eq:Action of SU(2) charged particle}
\end{equation}
where $Q^{a}=Tr(g\sigma_{3}g^{-1}t^{a})$, $t^{a}= {\half} \sigma^a$. 

The last term in (\ref{eq:Action of SU(2) charged particle}) is the
co-adjoint orbit action which describes the dynamics of the gauge
charges and which, upon quantization, gives the Hilbert space corresponding
to one unitary irreducible representation ($UIR$) of $SU(2)$ corresponding
to the highest weight $w/2$, hence of dimension $w+1$. $Q^{a}$
then become operators realizing the charge algebra
\begin{equation}
\left[Q^{a},Q^{b}\right]=if^{abc}Q^{c}.\label{eq: charge algebra}
\end{equation}
Under $g\rightarrow g\exp\left(i \sigma_3\phi /2\right)$,
the change in the action is given by $\Delta S=w\Delta\phi$. Thus single-valuedness
of $e^{iS}$ when $\phi$ traces out a closed path in $SU(2)$ leads
to the quantization of $w$. The crucial co-adjoint orbit term, when
generalized to several particles, becomes
\begin{equation}
S=-i\int dt\sum_{\lambda}w_{\lambda}\, \Tr\left(\sigma_{3}g_{\lambda}^{-1}\dot{g}_{\lambda}\right)\label{eq:multiparticle coadjoint orbit action}
\end{equation}
where we have a separate $g$ for each $\lambda$, and likewise for
$w$, with $\lambda$ indexing the particles. The continuum limit
of (\ref{eq:multiparticle coadjoint orbit action}) may be taken,
as one does for the Lagrange approach to fluids, by $\lambda\rightarrow{\bf x}$,
$\sum_{\lambda}\rightarrow\int {d^{3}{\bf x}/ v}$, ${w_{\lambda} /v}\rightarrow\rho({\bf x})$.
This leads to
\begin{equation}
S=-i\int d^{4}x~\rho \, \Tr(\sigma_{3}g^{-1}\dot{g})\label{eq: continuum form of Lagrange fluid}
\end{equation}
where $g=g({\bf x},t)$. Taking this as the leading term, namely, as the term
responsible for the symplectic structure, we can write an action
\begin{equation}
S=-i\int d^{4}x~j^{\mu}\, \Tr\left(\sigma_{3}g^{-1}D_{\mu}g\right)- \int F(n)+S_{YM}\label{eq:SU(2) gauge fluid}
\end{equation}
where $D_{\mu}g=\partial_{\mu}g+A_{\mu}g$, $A_{\mu}=-it^{a}A_{\mu}^{a}$,
$t^{a}= \sigma^a /2$.

The velocity for the transport of the nonabelian charge can be introduced
via $j^{\mu}=n\,u^{\mu}$, $u^{2}=1$. The current which couples to
the $SU(2)$ gauge field $A_{\mu}^{a}$ is given by
\begin{equation}
J^{a\mu}= \Tr\left(\sigma_{3}\,g^{-1}t^{a}g\right)j^{\mu}=Q^{a}j^{\mu}\label{eq:Eckert form of current SU(2)}
\end{equation}
which is in the Eckart form \cite{eckart}.
Starting with the action, one can easily verify the following \cite{Bistrovic:2002jx, review}:
\null\vspace{-6pt}
\begin{enumerate}[itemsep =-.02in]
\item The equations of motion for (\ref{eq:SU(2) gauge fluid}) do give
the appropriate magnetohydrodynamics.
\item The canonical quantization of (\ref{eq:SU(2) gauge fluid}) leads
to the expected current algebra. In particular, one finds the equal-time
rules
\begin{equation}
\left[\rho^{a}({\bf x},t),\, \rho^{b}({\bf y},t)\right]=if^{abc}\rho^{c}({\bf x},t)\, \delta^{3}({\bf x}-{\bf y}).\label{eq:current algebra}
\end{equation}
\end{enumerate}
The charge density, considered as a matrix in the fundamental representation,
transforms as $\rho\rightarrow h^{-1}\,\rho\, h$, $h\in SU(2)$, $\rho=\rho^{a} t^{a}$.
We can thus pick a specific $SU(2)$ transformation $g$ which diagonalizes
$\rho$,
\begin{equation}
\rho=g\,\rho_{diag}\,g^{-1}\label{eq:local diagonal density}
\end{equation}
so that $\rho^{a}=n\,Tr\left(g\sigma_{3}g^{-1}t^{a}\right)$. This identifies
the dynamical variable $g({\bf x},t)$ as part of the charge density.
The eigenvalues of $\rho$ are gauge-invariant and represented by
$n$. Their flow is given by $u^{\mu}$.

For a general gauge group $G,$ the action is given by
\begin{equation}
S=-i \int \sum_{s} \, j_{s}^{\mu}\, \Tr\left(q_{s}g^{-1}D_{\mu}g\right)- \int F(n_{1},n_{2},\ldots)+S_{YM}(A)\label{eq:general nonabelian charged gauge fluid action}
\end{equation}
where $q_{s}$ are the diagonal generators of $G$ and $j_{s}^{\mu}j_{s\mu}=n_{s}^{2},$
$s=1,2,\ldots,rank(G)$.

\section{Fluids and gravity}

We now go back to the case of the fluid with no nonabelian internal degrees of freedom.
As noted before, this case is described by the action (\ref{eq:group theoretic fluid action}) (or its relativistic version
(\ref{eq:group theoretic action for relativistic fluid})). Nevertheless, there is something not completely satisfactory about this. The group element $g$ belongs to $SU(1,1)$ and this group has no particular meaning in the relativistic theory.
We would like to analyze the effect of gravitational or mixed anomalies on the fluid equations.
The anomalies, as is well known, can be formulated in terms of the Lorentz group which acts on the tangent space or in terms of diffeomorphisms.
The former point of view requires identifying a Lorentz group action, while the latter can be related to Poincar\'e group action. The $SU(1,1)$ description does not immediately lead to an easily identifiable action of the Lorentz or Poincar\'e groups.
For this reason, we seek a generalization of the action
(\ref{eq:group theoretic action for relativistic fluid}); the particles underlying the fluid description may or may not have spin.

\subsection{Fluids with spin}

We start by considering an action similar to (\ref{eq:group theoretic action for relativistic fluid}) but with the group element
$g \in SU(1,1)$ replaced by an element of the Lorentz group, say,
$\Lambda$, in some finite dimensional matrix representation \cite{bal}.
The appropriate action is,
\beqar
S[e, \omega, j, \Lambda] &=&\int \det{e}\, \left[-i\, j^{\mu}\,\Tr (S_{12}\,\Lambda^{-1}D_{\mu}(\bar{\omega}(e))\,\Lambda)-F(n)\right]\nonumber\\
&&\hskip .2in -{1\over 32 \, \pi \, G} \epsilon_{abcd} \int e^a\wedge e^b \wedge R^{cd}(\omega).\label{eq: group action with gravity}
\eeqar
We have added the Einstein-Hilbert action for gravity as well, written in terms of the frame field one-form $e^a = e^a_\mu \, dx^\mu$ and the spin connection
$\omega^{ab} = \omega^{ab}_\mu \, dx^\mu$. $R^{ab}$ is the 
curvature two-form given by
\beqar
R^{ab} &=& d\, \omega^{ab}+\omega ^{ac} \wedge\omega^{cb} = {1\over 2} 
R^{ab}_{\mu\nu} \, dx^\mu \wedge dx^\nu\nonumber\\
&=& {1\over 2}\, e^a_{\,\alpha} \,(e^{-1})^{b \beta} \, (R_{\mu\nu})^\alpha_{~\beta}\, dx^\mu \wedge
dx^\nu.\label{curvature}
\eeqar
$\bar{\omega}(e)$ is the torsion free spin connection derived entirely from the metric or equivalently the frame fields. It is taken as understood that the
contraction of the tangent space indices is done with the flat Minkowski metric
$\eta^{ab}$. Coordinate indices are contracted, as needed, using the metric
$g_{\mu\nu} = e^a_\mu e^b_\nu \, \eta_{ab}$, so that
$n^2=j^{\mu}j^{\nu}g_{\mu\nu}$.
Further, in (\ref{eq: group action with gravity}), $S_{12}$ is a matrix corresponding to
the third component of the spatial spin, i.e., equal to
the corresponding Lorentz generator in the representation corresponding to
$\Lambda$.
By considering the right translations of $\Lambda$ by an element of the form
$\exp (i S_{12}\, \theta_{12} )$, we can see that
$j^{\mu}$ is a covariantly conserved current. Under the local
Lorentz transformation $\Lambda$, the transformation rules for the various quantities are
as follows.
\beqar
e^a\rightarrow e'\,^{a}=\Lambda^a_{~b}\,  e^b, &&\hskip .1in
\omega^a_{~b}\rightarrow\omega'\,^a_{~b} =
\Lambda^a_c\, \omega^c_{~d}\, (\Lambda^{-1})^d_b
- (d \,\Lambda \,\Lambda^{-1})^a_{~b},\nonumber\\
&&R^a_{~b} = \Lambda^a_{~c}\, R^c_{~d}\, (\Lambda^{-1})^d_{~b}.
\label{transform1}
\eeqar
The variation of the action (\ref{eq: group action with gravity}) with respect to
the spin connection $\omega$ gives the torsion free condition,
\beq
D\wedge e=0. \label{torsion}
\eeq
This can be solved to determine $\omega$ as a function of
$e$; we denote the solution as
$\bar{\omega}(e)$.
It corresponds to the spin connection derived from the metric
via the Christoffel symbols and is explicitly given by $\omega_\mu =
-i \,\omega^{ab}_\mu \, S_{ab}$ with
\beq
\bar{\omega}_{\mu}^{ab}=(e^{-1})^{\nu a}\partial_{[\mu}e_{\nu]}^{b}- (e^{-1})^{\nu b}\partial_{[\mu}e_{\nu]}^{a}- (e^{-1})^{\rho a}\,(e^{-1})^{\sigma b}\partial_{[\rho}e_{\sigma]}^{c}e_{\mu c}.
\label{omega-bar}
\eeq
In the action (\ref{eq: group action with gravity}), $\omega$ occurs only in
the last term; in the covariant derivative for $\Lambda$ we use ${\bar \omega}$ directly,
so that
\beq
\left(\Lambda^{-1}D_{\mu}\Lambda\right)^{a}\,_{b}=\left(\Lambda^{-1}\partial_{\mu}\Lambda+\Lambda^{-1}\, \bar{\omega}_{\mu}\, \Lambda \right)^{a}\,_{b}.
\label{cov-der}
\eeq
If we had used $\omega$ in this term, the condition for vanishing torsion, namely,
equation (\ref{torsion}), would be altered. The use of the solution
${\bar \omega}$ is similar to what is done for coupling
gravity to spin-${\half}$ particles, preserving the Riemannian or torsion-free condition.

In addition to the equation for $\omega$, there are equations of motion for $\Lambda$,
$j^\mu$ and $e^a_\mu$. The last one corresponds to the field equations for gravity.
For the variation of $\Lambda$, we can use
\begin{eqnarray*}
\delta\left(\Lambda^{-1}D_{\mu}\Lambda\right) & = & -\Lambda^{-1}\delta\Lambda\Lambda^{-1}D_{\mu}\Lambda+\Lambda^{-1}D_{\mu}(\delta\Lambda)\\
 & = & \Lambda^{-1}\partial_{\mu}(\delta\Lambda\Lambda^{-1})\Lambda+\Lambda^{-1}\left(\omega_{\mu}\delta\Lambda\Lambda^{-1}-\delta\Lambda\Lambda^{-1}\omega_{\mu}\right)\Lambda\\
 & = & \Lambda^{-1} (D_{\mu} \Theta )\, \Lambda.
\end{eqnarray*}
where $\Theta = \left(\delta\Lambda\,\Lambda^{-1}\right)$.
This leads to the equation of motion,
\begin{equation}
{1\over \sqrt{g}}\,D_{\mu}(\sqrt{g}\, j^{\mu}Q^{ab}) =   0\label{eq:variation wrt group element}
\end{equation}
where $Q^{ab}= \Tr\left(S_{12}\,\Lambda^{-1}\, S^{ab}\, \Lambda\right)$ is the spin density. Notice that the derivative involved in this divergence is Levi-Civita covariant  and also covariant
with respect to the Lorentz group action on the tangent space.
Similarly, right translations of $\Lambda$ in the $S_{12}$-direction gives
\beq
{1\over \sqrt{g}}\,\partial_{\mu}\left(\sqrt{g}\, j^\mu \right) \equiv \nabla_\mu \, j^\mu = 0.
\label{div-j}
\eeq
The equation of motion for $j^\mu$ becomes
\beq
j_\mu = - {n \over F'}\, i \, \Tr ( S_{12} \, \Lambda^{-1} \, D_\mu \Lambda ).
\label{j-eqn}
\eeq

The variation of the action with respect to the metric $g_{\mu\nu}$ (or equivalently,
the frame fields $e^a_\mu$) gives the standard terms except for the
the variation due to ${\bar \omega}$. The result is
\beqar
\delta S &=&  {1\over 2} \int  \sqrt{g}~ \delta g^{\mu\nu}\,\left[ 
T^{(f)}_{\mu\nu} - {1\over 8 \pi\, G}
(R_{\mu\nu} - {1\over 2} g_{\mu\nu} \, R)\right] ~+~\delta S_{extra}\nonumber\\
T^{(f)}_{\mu\nu} &=& n F' \, u_\mu u_\nu - g_{\mu\nu} (n F' - F ) \nonumber\\
\delta S_{extra} &=& -  \int \sqrt{g}\, j^\mu\,Q^{ab} \, \delta {\bar \omega}_{\mu ab}
\label{vary-g1}
\eeqar
where we have used (\ref{j-eqn}).
The last term can be simplified using
\beqar
\left(\delta{\bar\omega}_{\mu}\right)_{ab} & = & (e^{-1})^{\alpha}_{\,b} \left(\nabla_{\alpha}\delta e_{\mu,a}-\nabla_{\mu}\delta e_{\alpha,a}\right)
 - (e^{-1})^{\alpha}_{\, a} 
 \left(\nabla_{\alpha}\delta e_{\mu,b}-\nabla_{\mu}\delta e_{\alpha,b}\right)\nonumber\\
 && - (e^{-1})^{\alpha}_{\, a}\, (e^{-1})^{\beta}_{\, b}\,\left(\nabla_{\alpha}\delta e_{\beta}-\nabla_{\beta}\delta e_{\alpha}\right)^{m}e_{\mu}^{n}\, \eta_{mn.}
 \label{vary-g2}
\eeqar
Here $\nabla$ denotes the derivative covariant with respect to the tangent space and 
the Levi-Civita connection. With partial integrations and using
(\ref{eq:variation wrt group element}), this can be simplified as
\beq
\delta S_{extra} = \int \sqrt{g}~ \delta g^{\mu\nu}\, \nabla_\alpha (j_\mu Q^\alpha_{~ \nu}
+ j_\nu Q^\alpha_{~ \mu}) \label{vary-g3}
\eeq
where $Q^{\alpha\beta} = Q^{ab} (e^{-1})^\alpha_{\, a} (e^{-1})^\beta_{\, b}$.
Thus the energy-momentum tensor is not quite of the perfect fluid form
$T^{(f)}_{\mu\nu}$, rather it is
\beq
T_{\mu\nu} = T^{(f)}_{\mu\nu} ~+~ 2\, \nabla_\alpha ( j_\mu Q^\alpha_{~\nu} +
j_\nu Q^\alpha_{~\mu}).
\label{vary-g4}
\eeq
The conservation law becomes
\beq
\nabla_\mu \, T^{(f)\mu\nu} -
2\, (R_{\alpha\beta})^\nu_{~\lambda} \, j^\lambda Q^{\alpha\beta} = 0
\label{vary-g5}
\eeq
where we have used (\ref{eq:variation wrt group element}) and identities on the Riemann tensor.
The fluid has a spin density and the extra term in (\ref{vary-g5}) is very reminiscent of the
coupling of spin and curvature which appears at the point-particle level in the Mathisson-Papapetrou equations \cite{Mathisson:1937zz}. We may regard (\ref{vary-g5}) as the fluid generalization of
the latter.

\subsection{Spinless fluids}

As mentioned before, the basic reason for the description given above in terms of the Lorentz group is to incorporate
easily the gravitational anomalies
in the fluid language.
Since such anomalies, when they occur, 
are due to fields with spin, we may regard the action (\ref{eq: group action with gravity})
as adequate for our needs. Nevertheless, 
it is interesting at this point to consider an action for a fluid of spinless particles
so that the energy-momentum tensor has no extra term depending on the spin density.
Notice also that, from (\ref{j-eqn}), it is the transport of spin
which is described by the current $j^\mu$
and not particle number or mass; in other words, we have a fluid of spin carriers, spin being their only attribute.
It would be useful to consider the flow arising from transport of mass.
The natural object for this would be the Poincar\'e group, in particular, the translations, since spinless particles do have transport of energy and momentum.
Since we will need matrix representations and traces, it is easier to consider
the Poincar\'e group as obtained from the de Sitter group $SO(4,1)$ via a group contraction.
In addition to the Lorentz generators $S_{ab}$, we then have
$P_a = S_{a5}/ R$ which are the translations (on de Sitter space) with
\beq
[P_{a},P_{b}] = i\, \frac{S_{ab}}{R^{2}}. \label{Poin1}
\eeq
The limit $R \rightarrow \infty$ corresponds to the group contraction and reduces the algebra to the Poincar\'e algebra. A specific matrix realization of the $SO(4,1)$-algebra is provided by
the Dirac $\gamma$-matrices $\gamma_{ab}$ and $\gamma_a \gamma_5$.

If $g$ denotes an element of $SO(4,1)$, then the frame fields for the coset space are given by
$e^a =  -i \Tr ( S^{a5} g^{-1} \, d g)$, and the metric is given by
\beq
ds^{2}= - \Tr(S^{a5}g^{-1}dg)\,\Tr(\S_{a5} g^{-1}dg).\label{Poin2}
\eeq
The action for a point-particle is thus
\beqar
I[g] &=& - m \int dt\sqrt{-\Tr (S^{a5}g^{-1}\dot{g})\,\Tr(S_{a5}g^{-1}\dot{g})}\nonumber\\
&=& - {1\over 2} \int dt~\left[  m^2 \, \eta -  {(\Tr (S^{a5}g^{-1} {\dot g} ))^2 \over \eta}\right]
\label{Poin3}
\eeqar
where, in the second line, we have used a world-line metric as an auxiliary field.
We can further reduce this as
\beq
I[g] = \int dt~ \left[  j^a (-i\, \Tr (S_{a5} \,g^{-1} {\dot g}))  + {\eta\, (j^a\, j_a -m^2) \over 2}.
\label{Poin4}
\right]
\eeq
The similarity with the fluid actions we have discussed is evident. This suggest that, for spinless fluids, we can use the action
\beq
S= 
\int d^{4}x\sqrt{g}\,\left[-i\,j^{a\mu}\,\Tr\left(S_{a5}\,g^{-1}\partial_{\mu}g\right)
~-~ F(n)\right].\label{Poin5}
\eeq
Coupling to gravity is introduced by $\del_\mu g \rightarrow D_\mu g
= (\del_\mu + {\bar \omega}_\mu ) g$, where ${\bar \omega}$ is the torsion-free spin connection as before. The full action is thus
\beq
S =\int \det e\,\left[-i\,j^{a\mu}\,\Tr\left(S_{a5}\,g^{-1}\,D_{\mu}g\right)
- F(n)\right]  - {1\over 32 \pi \,G}\epsilon_{abcd} \int e^a\wedge e^b \wedge R^{cd}(\omega).
\label{Poin6}
\eeq

The derivation of the equations of motion will proceed as before. The terms involving
$g$ will give the energy-momentum tensor of the perfect fluid form, except for the
term coming from the variation of ${\bar \omega}$; in other words,
\beq
T_{\mu\nu} = T^{(f)}_{\mu\nu} ~+~ 2\, \nabla_\alpha ( j^a_\mu Q^\alpha_{~\nu a} +
j^a_\nu Q^\alpha_{~\mu a})
\label{Poin7}
\eeq
where $Q^{\alpha \beta}_c=  \Tr (g\,S_{c5} \,g^{-1}  S^{ab}) (e^{-1})^\alpha_a (e^{-1})^\beta_b$.
Since $S_{c5}$ and $S_{ab}$ are orthogonal with the trace,
$Q^{\alpha\beta}_c$ vanishes unless $g \, S_{c5} \, g^{-1}$ generates
a term proportional to the Lorentz generator $S_{ab}$. This can only be done via the
commutator $[S_{c5}, S_{d5}]$ for terms in $g$ which are of the form
$\exp (i S_{d5}\, \theta^{d5} + \cdots)$. As a result, $Q^{\alpha\beta}_c$
is of order $1/R^2$ and vanishes in the contraction limit.  The energy-momentum tensor then has the perfect fluid form.
Thus, the action
(\ref{Poin6}) can describe spinless fluids in terms of the Poincar\'e group realized as the contraction limit of the de Sitter group.

\section{Standard Model}

We are now in a position to apply this to the standard model and a
fluid or plasma phase of the same. For specificity consider the quark-gluon
plasma phase for three flavors of quarks, $u,d,s$. In other words,
we consider a phase with thermalized $u,d,s$ quarks, so that they
must be described by fluid variables while the heavier quarks are
described by the field corresponding to each species. The flavor symmetries,
for the moment, will be taken to be gauged. We will also neglect the
quark masses so that we have the full flavor symmetry $U(3)_{L}\times U(3)_{R}$.
Thus the group $G$ to be used in (\ref{eq:general nonabelian charged gauge fluid action})
is 
\begin{equation}
G=SU(3)_{c}\times U(3)_{L}\times U(3)_{R}\label{eq:standard model color and flavor}
\end{equation}
with individual flows corresponding to the charges. In this discussion
our focus is on the flavor transport, so we will drop $SU(3)_{c}$
from the equations to follow. Of course, the flavor symmetry is not
fully preserved even in the absence of masses due to anomalies. On
this question, we then have a rerun of the old 't Hooft argument \cite{'tHooft:1979bh}.

Assume all flavor symmetries are gauged with anomalies canceled by
a set of spectator fermions. In the fluid phase where $u,d,s$ are
replaced by fluid variables, we must then have a term in the fluid
action which can reproduce the anomalies so that the cancellation
with spectator fermions still remains valid. (In the usual case where
the phase being considered is one of confinement and chiral symmetry
breaking, this term is the Wess-Zumino term constructed in terms of
the pseudoscalar meson fields.). In the present case, since we already
have a description of the fluid in terms of group elements, it is
easy enough to adapt the usual Wess-Zumino term. Thus our fluid action
is given by
\begin{eqnarray}
S & = & \int \biggl[ -i\, j_{3}^{\mu}\Tr\left(\frac{\lambda_{3}}{2}g_{L}^{-1}D_{\mu}g_{L}\right) -i\, j_{8}^{\mu}\Tr\left(\frac{\lambda_{8}}{2}g_{L}^{-1}D_{\mu}g_{L}\right)
 -i\, j_{0}^{\mu}\Tr\left(g_{L}^{-1}D_{\mu}g_{L}\right)
 \nonumber \\
 & & \hskip .25in -i\,k_{3}^{\mu}\Tr\left(\frac{\lambda_{3}}{2}g_{R}^{-1}D_{\mu}g_{R}\right) - i\,k_{8}^{\mu}\Tr\left(\frac{\lambda_{8}}{2}g_{R}^{-1}D_{\mu}g_{R}\right)
-i\, k_{0}^{\mu}\Tr\left(g_{R}^{-1}D_{\mu}g_{R}\right)\nonumber \\
&& \hskip .25in -F(n_{3},n_{8},m_{3}m_{8}) +S_{YM}(A)\nonumber \\
&& \hskip .25in+\,\Gamma_{WZ}(A_{L},A_{R},g_{L}g_{R}^{\dagger})\biggr]\label{eq:complete action}
\end{eqnarray}
where $j_{0,3,8}^{\mu}$ apply to $U(3)_{L}$ and $k_{0,3,8}^{\mu}$
apply to $U(3)_{R}$ and $g_L \in U(3)_L$, $g_R \in U(3)_R$.
The last term is the usual gauged WZ term $\Gamma_{WZ}(A_{L},A_{R},U)$
given in terms of $A_{L},A_{R}$ and the meson fields $U\in U(3)$,
and gauged
in a way that preserves the vector symmetries,
but, for our purpose, $U$ is replaced by $g_{L}g_{R}^{\dagger}$. Explicitly $\Gamma_{WZ}$
is given by Witten in \cite{Witten:1983tw} as
\begin{eqnarray}
\Gamma_{WZ} & = & -\frac{iN}{240\pi^{2}}\int \Tr\left(dUU^{-1}\right)^{5}\nonumber \\
 &  & +\frac{iN}{48\pi^{2}}\int \Tr\left(A_{L}dA_{L}+dA_{L}A_{L}+A_{L}^{3}\right)dUU^{-1}\nonumber \\
 &  & +\frac{iN}{48\pi^{2}}\int \Tr\left(A_{R}dA_{R}+dA_{R}A_{R}+A_{R}^{3}\right)U^{-1}dU\nonumber \\
 &  & -\frac{iN}{96\pi^{2}}\int \Tr\left[\left(A_{L}dUU^{-1}\right)^{2}-\left(A_{R}U^{-1}dU\right)^{2}\right]\nonumber \\
 &  & -\frac{iN}{48\pi^{2}}\int \Tr\left[A_{L}\left(dUU^{-1}\right)^{3} + A_{R}\left(U^{-1}dU\right)^{3}\right]\nonumber \\
 &  & -\frac{iN}{48\pi^{2}}\int \Tr\left(dA_{L}dUA_{R}U^{-1}-dA_{R}dU^{-1}A_{L}U\right)\nonumber \\
 &  & -\frac{iN}{48\pi^{2}}\int \Tr\left(A_{R}U^{-1}A_{L}U(U^{-1}dU)^{2}-A_{L}UA_{R}U^{-1}(dUU^{-1})^{2}\right)\nonumber \\
 &  & +\frac{iN}{48\pi^{2}}\int \Tr\left((dA_{R}A_{R}+A_{R}dA_{R})U^{-1}A_{L}U-(dA_{L}A_{L}+A_{L}dA_{L})UA_{R}U^{-1}\right)\nonumber \\
 &  & +\frac{iN}{48\pi^{2}}\int \Tr\left(A_{L}UA_{R}U^{-1}A_{L}dUU^{-1}+A_{R}U^{-1}A_{L}UA_{R}U^{-1}dU\right)\nonumber \\
 &  & -\frac{iN}{48\pi^{2}}\int \Tr\left(A_{R}^{3}U^{-1}A_{L}U-A_{L}^{3}UA_{R}U^{-1}+\frac{1}{2}UA_{R}U^{-1}A_{L}UA_{R}U^{-1}A_{L}\right). \label{eq:Covariant form of Wess-Zumino}
\end{eqnarray}
One of the main results of this paper is that the action given above incorporates
 all the flavor anomalies in fluid dynamics.
Once we have obtained (\ref{eq:Covariant form of Wess-Zumino}), we
can restrict the gauge fields $A_{L},A_{R}$ to what is needed for
the standard model, namely the $SU(2)\times U(1)$ group of electroweak
interactions. It is straightforward to verify that (\ref{eq:Covariant form of Wess-Zumino})
does indeed lead to the usual chiral magnetic effect.

\section{Currents from anomalies}

\subsection{The chiral magnetic and chiral vorticity effects}

For the chiral magnetic effect, we have only a background electromagnetic
field turned on, so that
$A_{L}=A_{R}= -i\,Q\,A $, 
where $Q$ is the quark charge matrix given by
$Q = diag ({2\over 3}, - {1\over 3}, -{1\over 3})$.
The contribution of the anomaly to the electromagnetic current 
following from (\ref{eq:Covariant form of Wess-Zumino}) has been given in
\cite{callan}.
The electromagnetic current for the action
(\ref{eq:complete action}), (\ref{eq:Covariant form of Wess-Zumino}) is then given by
\beqar
J^\mu&=& J_3^{\mu} + {e\over 16\pi^2} \epsilon^{\mu\nu\alpha\beta}
\Tr \left[ Q (\del_\nu U \, U^{-1}\, \del_\alpha U \, U^{-1}\,\del_\beta U \, U^{-1})
+(U^{-1} \del_\nu U\, U^{-1} \del_\alpha U\, U^{-1} \del_\beta U) \right]\nonumber\\
&&+ i {e^2 \over  4\pi^2}  \epsilon^{\mu\nu\alpha\beta}
\del_\nu A_\alpha \Tr \left[ Q^2 (\del_\beta U \, U^{-1} + U^{-1} \del_\beta U)
+{1\over 2} (Q \del_\beta U \, Q U^{-1} - Q U Q \del_\beta U^{-1})
\right]\nonumber\\
\label{chimag1}
\eeqar
where $J_3^{\mu}$ is the contribution from the nonanomalous part of the action
and we have set $N =3$.
For the main points we want to illustrate, it is sufficient to consider
a reduction to the $SU(2)$ subgroup; in other words, we will consider
basically the up and down quarks. In this case, we can take
\beq
U = e^{i\theta}\, \left[ \begin{matrix}
V&0\\
0&1\\
\end{matrix}\right]
\label{chimag2}
\eeq
where $V$ is a $2\times 2$ matrix which is an element of $SU(2)$.
We take it to be of the form
$V = g_L \, g_R^\dagger$, where $g_L$ and $g_R$ are now elements of $SU(2)$.
The nonanomalous part of the current can then be written as
\beq
J_3^{\mu} = -{1\over 4} \left[ n_3 \, u^\mu_{3L} \Tr ( \sigma_3 \, g_L^{-1} \sigma_3 g_L)
+ m_3\, u^\mu_{3R} \Tr (\sigma_3 g_R^{-1} \sigma_3 g_R)\right]
\label{chimag3}
\eeq
Simplifying (\ref{chimag1}) with the choice of $U$ we have made,
\beqar
J^\mu &=& J_3^{\mu} +
{e \over 48 \pi^2} \epsilon^{\mu\nu\alpha\beta} \Tr ( {\cal I}_\nu \, {\cal I}_\alpha \,{\cal I}_\beta)
+ i {e^2 \over 16 \pi^2} \epsilon^{\mu\nu\alpha\beta}\, \del_\nu A_\alpha\,
\Tr\left[ ( \Sigma_{3L} + \Sigma_{3R} ) \, I_\beta \right] ~+ J^\mu_\theta\nonumber\\
J^\mu_\theta &=& - {e^2 \over 4 \pi^2} \epsilon^{\mu\nu\alpha\beta}\, \del_\nu A_\alpha\,\del_\beta \theta\,
\left[ 2 + {1\over 4} \Tr \left(\Sigma_{3L} \,\Sigma_{3R} - 1\right)\right]
\label{chimag4}
\eeqar
where ${\cal I}_\beta = g_L^{-1}\del_\beta g_L - g_R^{-1} \del_\beta g_R$ and
$\Sigma_{3L} = g_L^{-1} \sigma_3 g_L$, $\Sigma_{3R} = g_R^{-1} \sigma_3 g_R$.
In simplifying (\ref{chimag1}) to this form, we have used the fact that
there is no rank-3 symmetric invariant tensor for $SU(2)$.

When $g_L = g_R$, the last term in (\ref{chimag4}), namely, $J^\mu_\theta$, reduces to
\beq
J^\mu_\theta = - {e^2 \over 2 \pi^2} \epsilon^{\mu\nu\alpha\beta} (\del_\nu A_\alpha) \, \del_\beta \theta. \label{chimag5}
\eeq
This is the chiral magnetic effect discussed in \cite{Kharzeev:2004ey}.
The quantity $\nabla \theta$ is related to the fluid current for the transport of the $U(1)_A$ axial
charge. And, correspondingly, in a medium in equilibrium, with chiral asymmetry for such charges,
we may replace $ {\dot \theta}$ by ${\half} (\mu_L - \mu_R)$, where the chemical potentials
are for the left and right axial charges. Notice, however, that the expression for
$J^\mu_\theta$ has added terms when $g_L$ are $g_R$ are independent matrices.

The other terms in (\ref{chimag4}) can be simplified further.
First of all, using the Maurer-Cartan equations
$d (g^{-1} d g) + (g^{-1} dg )^2 = 0$, we can simplify
\beq
\epsilon^{\mu\nu\alpha\beta} \,\Tr (g^{-1} \del_\nu g\,g^{-1} \del_\alpha g\, g^{-1} \del_\beta g)
=  i  \Tr (\sigma_3 \, g^{-1} \del_\nu g )~ \del_\alpha \left[ i \Tr ( \sigma_3 g^{-1} \del_\beta g)\right].
\label{chimag6}
\eeq
We can use this to simplify the term $\epsilon^{\mu\nu\alpha\beta} \Tr ( {\cal I}_\nu \, {\cal I}_\alpha \,{\cal I}_\beta)$ in (\ref{chimag4}). Further, from the equation of motion for
$j^\mu_3$ and $k^\mu_3$, we find
\beqar
i \, \Tr (\sigma_3 \,g_L^{-1} \del^\mu g_L ) &=& - 2 \, {\del F \over \del n_3} \, u_{3L}^\mu
= - {2 \over n_3} {\del F \over \del n_3} \, j^\mu_{3}\nonumber\\
i \, \Tr (\sigma_3 \, g_R^{-1} \del^\mu g_R ) &=& - 2 \, {\del F \over \del m_3} \, u_{3R}^\mu
= - {2 \over m_3} {\del F \over \del m_3} \, k^\mu_{3}
\label{chimag7}
\eeqar
where $u_{3L}^\mu$ and $u_{3R}^\mu$ are the flow velocities for the left and right
isospin.
Using these results, the current finally takes the form
\beqar
J^\mu &=& J_3^{\mu} + J^\mu_\theta  + i {e^2 \over 16 \pi^2} \epsilon^{\mu\nu\alpha\beta}\, \del_\nu A_\alpha\,
\Tr\left[ ( \Sigma_{3L} + \Sigma_{3R} ) \, I_\beta \right]\nonumber\\
&&\hskip .2in +{1\over 16 \pi^2} \epsilon^{\mu\nu\alpha\beta}
\del_\nu \Tr ( g_L^{-1} \del_\alpha g_L\,\, g_R^{-1} \del_\beta g_R)\nonumber\\
&&\hskip .2in + {e \over 12 \pi^2} \epsilon^{\mu\nu\alpha\beta} \left[ \left( {\del F \over \del n_3}\right)^2
\, u_{3L\, \nu} \, \del_\alpha u_{3L\, \beta} -
 \left( {\del F \over \del m_3}\right)^2
\, u_{3R\, \nu} \, \del_\alpha u_{3R\, \beta} \right].
\label{chimag8}
\eeqar
The last term of this expression involves the vorticity of the flow velocities.
This equation is thus an expression of the chiral vorticity effect.

\subsection{Mixed gauge-gravity anomaly}

In addition to the flavor anomalies, it is also possible to consider the mixed gauge-gravity anomaly in the standard model. The six-form index density which leads to this via the descent equations
is
\beq
I_{6}= {i \over 384\,\pi^3}\,\left(\Tr F\right)\, \Tr\left(R\wedge R\right)
\eeq
where the field strength is the one corresponding to the weak hypercharge
$U(1)_Y$. The trace of the hypercharge vanishes for each generation of quarks
by itself, so that this anomaly is zero. The possibility of a contribution arises when we consider
a plasma where some of the quarks, say, the up, down and strange quarks, are in the fluid phase while others, say, charm, is to be described by the standard fermion Lagrangian.
In this case, for the fluid part we would need an effective description.

There are two choices on how this anomaly can be displayed; we can choose to regard this as an anomaly in the hypercharge current or as an anomaly in local Lorentz transformations.
For the first point of view, the index density leads, via the descent equations to the 
effective action
\beq
\Gamma_{WZ} = i {N\over 192 \pi^2} \,
\int \Tr ( d \theta ) \, \Tr \left( \omega \, d\omega + {2\over 3} \omega^3\right).
\label{grav1}
\eeq
The hypercharge current has the conservation law
\beq
\del_\mu J^\mu = - i{N \over 768\,\pi^2} \, {\epsilon^{\mu\nu\alpha\beta} \over \sqrt{g}}\,
\Tr (R_{\mu\nu} \, R_{\alpha\beta}).
\label{grav1a}
\eeq
Further, if we choose to regard $\omega$ as an independent quantity, then the torsion-free condition is modified by the Lorentz Chern-Simons term, when this term is added to the Einstein-Hilbert action.
The more canonical thing to do would be to consider $\omega$ in 
(\ref{grav1}) to be the solution ${\bar \omega}$. In this case, with 
$\omega \rightarrow {\bar \omega}$, we find for the correction to the energy-momentum tensor,
\beq
T^{\nu \sigma}\bigr]_{corr}
= - i {N \over 192 \pi^2} \, {1\over \sqrt{g}}\nabla_\lambda \left[ \Tr (\del_\mu \theta)\, (R_{\alpha\beta})^{\lambda \sigma}
\epsilon^{\mu\nu\alpha\beta} + ({\nu \leftrightarrow \sigma}) \right].
\label{grav2}
\eeq
The remaining trace is over the hypercharge values. If we replace
${\dot \theta}$ by the chemical potentials, as can be done for the chiral magnetic effect,
\beq
\Tr ( {\dot \theta} ) \rightarrow {1\over 2} \left[ {1\over 3} ( \mu^u_L + \mu^d_L + \mu^s_L)
+{2\over 3} ( \mu^d_R + \mu^s_R - 2 \mu^u_R)\right].
\label{grav3}
\eeq
More generally, we can replace $\del_\mu \theta$ by its expression from the equation of motion
giving a term involving the derivative of the enthalpy function, similar to what was
done in (\ref{chimag8}). 
(Since the derivative of the enthalpy at fixed entropy and pressure is the chemical potential, 
this includes the previous case as well.) 
Thus, depending on the properties of the enthalpy function of the fluid,
 the corrections displayed in (\ref{grav2}) can be nonzero even when $\mu_i =0$.

The other possibility is to consider the index density as leading to anomalies in local Lorentz
transformations. We can use an element of the Lorentz group, identified as
the fluid variable $\Lambda$ of section 3, to write the Wess-Zumino term.
The transformation of fields of the relevant fields is given by
\beqar
e\rightarrow e^{g}&=&g\, e, \hskip .2in  \Lambda \rightarrow \Lambda^g = g \, \Lambda\nonumber\\
\omega\rightarrow\omega^{g} 
 & = & g\, \omega\, g^{-1} - dg\, g^{-1}, \hskip .2in R\rightarrow R^{g}=g\, \omega\,  g^{-1}.
 \label{grav4}
\eeqar
The Wess-Zumino term may then be written as
\beqar
\Gamma_{WZ}&=&  i {N \over 192\pi^2}\int  \Tr (F) \, \left[ \Tr \left(\omega \, d\omega + {2\over 3} \omega^3\right)
- \Tr \left(\Omega \, d\Omega + {2\over 3} \Omega^3\right) \right]\nonumber\\
&=&  i {N \over 192\pi^2} \int \left[\Tr (F) \, \Tr (d \Lambda \, \Lambda^{-1} \, { \omega})
+ {1\over 3} \Tr (F) \,\Tr (d \Lambda \, \Lambda^{-1} )^3\right]
\label{grav5}
\eeqar
where $\Omega = \Lambda^{-1} \left[ d \, \Lambda + \omega \, \Lambda\right]$.
Once again, if we regard $\omega$ as independent, then this leads to a nonzero torsion proportional to the spin-density.
The equation of motion for $\omega$, starting from
(\ref{eq: group action with gravity}) and adding (\ref{grav5}), can be reduced to the form
\beqar
{\epsilon^{\alpha\beta\mu\nu } \over 4 \pi\,G} \, (T_{\mu\nu})^a
&=& (M^\alpha)^{ab} \, (e^{-1})^\beta_{\, b} - (M^\beta)^{ab} \, (e^{-1})^\alpha_{\, b}
+ (M^\gamma)^{cd} \,(e^{-1})^\alpha_{\, c} (e^{-1})^\beta_{\, d}  \, e^a_\gamma
\nonumber\\
(M^\beta)^{cd} &=&
- {N \over 192\,\pi^2} \, \epsilon^{\mu\nu\alpha\beta} \epsilon^{abcd} \, \Tr (F_{\mu\nu})\,
\Tr (\del_\alpha \Lambda\, \Lambda^{-1} \, S_{ab} ) 
\label{grav6}
\eeqar
where $(T_{\mu\nu})^a = (D_\mu e_\nu)^a - (D_\nu e_\mu)^a$ is the torsion
tensor.

In the case when we use ${\bar \omega}$ in place of $\omega$ in (\ref{grav5}), we get corrections to the equation of motion.
For variations corresponding to the right translations of $\Lambda$ by a term proportional to $S_{12}$, we find
\beq
{1\over \sqrt{g}}\, D_\mu (\sqrt{g}\, j^\mu) \, \Tr(S_{12}^2)
= - {N\over 192 \pi^2}   \,  { \epsilon^{\mu\nu\alpha\beta}\over \sqrt{g}}\,\Tr (F_{\mu\nu})\,
\del_\alpha \Tr (S_{12}\, \Lambda^{-1}\, D_\beta \Lambda).
\label{grav7}
\eeq
The equation for the left translations of $\Lambda$ by an arbitrary infinitesimal Lorentz transformation is
\beq
{1\over \sqrt{g}}\, D_\mu \left[ \sqrt{g}\, j^\mu \, \Lambda S_{12} \Lambda^{-1}\right]
= {N \over 192\pi^2} \, { \epsilon^{\mu\nu\alpha\beta}\over \sqrt{g}}\,
\Tr (F_{\mu\nu})\, \left[ R_{\alpha \beta} - D_\alpha ( D_\beta \Lambda \, \Lambda^{-1})
\right].
\label{grav8}
\eeq
(Of course, the two equations, (\ref{grav7}) and (\ref{grav8}), are not completely independent.)

The variation of (\ref{grav5}) with respect to the frame field $e^a_\sigma$ will yield the correction to the energy-momentum tensor. This is given by
\beqar
T^\sigma_{~a}\bigr]_{corr} &=&
- {1\over \sqrt{g}}\, {i N \over 96\,\pi^2} \, 
e_{\lambda, a} \nabla_\beta\left[ \Tr (F_{\mu\nu}) \left( \Tr (\del_\alpha \Lambda\, \Lambda^{-1}
S^{\beta\sigma})\, \epsilon^{\mu\nu\alpha\lambda}
+  \Tr (\del_\alpha \Lambda\, \Lambda^{-1}
S^{\beta\lambda}) \,\epsilon^{\mu\nu\alpha\sigma} \right) \right]
\nonumber\\
&&\hskip .3in - {1\over \sqrt{g}}\, {i N \over 96\,\pi^2} \, 
e_{\lambda, a} \nabla_\beta\left[ \Tr (F_{\mu\nu}) \Tr (\del_\alpha \Lambda \, \Lambda^{-1} 
S^{\lambda \sigma} )\, \epsilon^{\mu\nu\alpha\beta}\right]
\label{grav9}
\eeqar
where $S^{\alpha\beta} = S^{ab} (e^{-1})^\alpha_{\,a} (e^{-1})^\beta_{\, b}$. The terms in the first line
of this equation leads to a symmetric energy-momentum tensor when written in terms of the coordinate components, by multiplying with $(e^{-1})^{\rho\, a}$. The term in the
second line leads to an antisymmetric term. This is to be expected. We know that a symmetric energy-momentum tensor is necessary for the conservation of the current corresponding to the Lorentz transformations. In the present case, the anomaly implies that this current is not conserved. The antisymmetric term is a manifestation of this property.
Since the Einstein tensor $R_{\mu\nu} - {1\over 2} g_{\mu\nu} R$ is symmetric, this leads to a problem with the Einstein equations. The proper way to understand this is to realize that there is another term in $T_{\mu\nu}$ due to the quarks we have neglected, say the charm quark
in the example we have been using. Since the latter field by itself also leads to a mixed anomaly, the full energy momentum tensor  which is the sum of
$T^{(f)}_{\mu\nu}$, $T_{\mu\nu}$ from (\ref{grav9}) and $T_{\mu\nu}$ from charm will together be symmetric, the anomaly part from the charm quark canceling the antisymmetric piece
of (\ref{grav9}).

\section{Discussion}

We have obtained a very general formalism for incorporating the effects of anomalies in hydrodynamics.  As mentioned in the introduction, being a formalism based on symmetries rather than calculations specific to any particular assembly of material particles, this is quite general and is expected to be valid beyond weak coupling or near equilibrium conditions. The specific choice of the fluid will be reflected in the choice of the enthalpy functions.

 A few clarifying comments are in order. 
 It is important to realise that in any fluid where the particles which constitute it carry a variety of quantum numbers, we can have a number of different flow velocities. This is evident from the action
 (\ref{eq:general nonabelian charged gauge fluid action}) where we have flow velocities for
 each diagonal generator of the group. This point seems not to be adequately
 emphasized in the literature.
 It is also useful to visualize this as follows. Consider two quarks and two antiquarks in a fluid. We could have them forming a color singlet and moving in the same direction. This gives a mass/energy flow but no color flow. We could visualize a  $q\, {\bar q}$ pair forming an octet state and moving together in a certain direction while the other $q\, {\bar q}$ pair form a singlet. This gives a nonzero color transport rate different from the mass/energy flow. We could also envisage subsets of particles forming different spin states giving a spin flow velocity, possibly different from the mass and color flows. When we consider massless quarks, the L, R quantum numbers are independent quantum numbers with independent velocities possible. 

Specifically for the flavor part,  we can have independent $u_{3L}$ and $u_{3R}$.
These need not coincide even when $g_L = g_R$ for two reasons: The local charge representation is determined by $n_3, m_3$ and these need not be the same even when $g_L = g_R$. Secondly, the enthalpies can be different as well. 

If a calculation is carried out in a specific medium, the results obtained would be for the appropriate enthalpy function. For example, if we take a massless field, then the relation between pressure and energy density corresponds to an enthalpy function $F \sim n^{4/3}$.
In this case, $(F')^2$  is of the form $n^{2/3}$. Notice that for the vorticity term 
in (\ref{chimag8})  there is a prefactor proportional to this. If, in addition, we take $n \sim T^3$, as is appropriate for a relativistic gas, then the prefactor gives a $T^2$ term. 
This may give a point of correspondence with the results in \cite{landsteiner}.
However, we should expect a contribution even at zero temperature, since the structure of the anomaly has to be reproduced correctly in the fluid language; this is evident from
section 5.2.
The derivative of the enthalpy function is also related
to the chemical potentials, when the latter is introduced.  Our formula
(\ref{chimag8}) is thus similar to the results in \cite{oz} as well.

Regarding the use of the Wess-Zumino term for anomalies, the specific choice of $\Gamma_{WZ}$ specifies the nature of the currents being discussed. (This point is moot for our discussion in sections 2 and 3, since we have not introduced anomalies yet.)
We have used the form (\ref{eq:Covariant form of Wess-Zumino}) which gives expressions invariant under the nonanomalous vector gauge symmetries.

The Wess-Zumino term was also used to obtain anomalies for
chiral superfluids in 
\cite{shulin}, although the formalism is very different from ours and
the emphasis was on baryonic and axial currents.
(This article came to our attention after this paper was completed. We thank the author
for correspondence on this.)
However, we may note  that equation (58) of 
\cite{shulin} is similar to our (\ref{chimag8}) if our $u_{3L}$ and $u_{3R}$ are related to
the different superfluid velocities introduced in that paper.

\vskip .1in
This work was supported by U.S.\ National Science
Foundation grant PHY-0855515
and by a PSC-CUNY grant.
%%%%%%%%%%%%%%%%%%%%%%%%%%%%%%%%%%%%%%%%%%%%%%%%
%%%%%%%%%%%%%%%%%%%%%%%%%%%%%%%%%%%%%%%%%%%%%%%%

%%%%%%%%%%%%%%%%%%%%%%%%%%%%%%%%%%%%%%%%%%%%%%%%
%%%%%%%%%%%%%%%%%%%%%%%%%%%%%%%%%%%%%%%%%%%%%%%%
%%%%%%%%%%%%%%%%%%%%%%%%%%%%%%%%%%%%%%%%%%%%%%%%
%%%%%%%%%%%%%%%%%%%%%%%%%%%%%%%%%%%%%%%%%%%%%%%%

\begin{thebibliography}{99}
\bibitem{bal} Actions similar to the one we are writing here, but for point-particles and strings, have been discussed by
A.P. Balachandran, F. Lizzi and G. Sparano, \NP~{\bf B277}, 359 (1986).
\end{thebibliography}
\end{document}